\documentclass[12pt]{article}

\usepackage{amsmath}
\usepackage{amssymb}
\usepackage{amsfonts}
\usepackage[mathscr]{euscript}

\title{\bf Determination of the noise parameters in a one-dimensional open quantum system}

\author{Alexandra M. Liguori$^{a,b}$, Giovanni Moras\\
\small $^a$Dipartimento di Fisica Teorica, Universit\`a di Trieste,
Strada Costiera 11,\\
\small 34014 Trieste, Italy\\
\small $^b$Istituto Nazionale di Fisica Nucleare, Sezione di
Trieste, 34100 Trieste, Italy \\
\small \texttt{liguori@ts.infn.it, morasg@gmail.com}}

\date{\null}

\begin{document}

\maketitle

\vskip 1cm

\begin{abstract}
\noindent
We consider an electron magnetically interacting with a spin-$1/2$ impurity,
embedded in an external environment whose noisy term acts only on the impurity's spin,
and we find expressions for the electron transmission and reflection probabilities in terms of the
phenomenological noise parameters. \\
Moreover, we give a simple example of the necessity of complete positivity for physical consistency,
showing that a positive but not completely positive dissipative map can lead to negative transmission
probabilities.
\end{abstract}

\section{Introduction}
In standard quantum mechanics the focus is mainly upon
\textit{closed} physical systems, i.e. systems which can be
considered isolated from the external environment and whose
reversible time-evolution is described by one-parameter groups of
unitary operators. On the other hand, when a system $S$ interacts
with an environment $E$, it must be considered as an \textit{open}
quantum system whose time-evolution is irreversible and exhibits
dissipative and noisy effects. A standard way of obtaining a
manageable dissipative time-evolution of the density matrix
$\varrho_t$ describing the state of $S$ at time $t$ is to construct
it as the solution of a Liouville-type Master equation $\partial_t
\varrho_t = \mathbf{L}[\varrho_t]$. This can be done by tracing away
the environment degrees of freedom and by
performing a Markovian approximation~\cite{lendi,verri,spohn}, i.e.
by studying the evolution on a slow time-scale and neglecting fast
decaying memory effects. Then the irreversible reduced dynamics of
$S$ is described by one-parameter semigroups of linear maps obtained
by exponentiating the generator $\mathbf{L}$ of Lindblad
type~\cite{kossak1, kossak2}:
$\Gamma_t = e^{t\mathbf{L}}$, $t \geq 0$, such that 
$\varrho_t \equiv \Gamma_t[\varrho]$.
By means of standard \textit{weak} or \textit{singular coupling limit} techniques, the Master equation
for $S$ can be rewritten as follows~\cite{gorini,dumcke,ben-flore}:
\begin{equation}
\label{master-eq} \frac{\partial \varrho_t}{\partial t} =
\mathbf{L}_H[\varrho_t] + \mathbf{L}_D[\varrho_t] =
-\frac{i}{\hbar}[H,\varrho_t] + \mathbf{L}_D[\varrho_t] \, ,
\end{equation}
where $\mathbf{L}_D[\varrho_t]$ is the Kossakowski-Lindblad term describing dissipation and noise.
If the system $S$ consists of a qubit, the Kossakowski-Lindblad term can be written explicitly as follows:
\begin{equation}
\label{dissipation} \mathbf{L}_D^{(2)}[\varrho(t)] = \sum_{i,j=1}^3
C_{ij} \Bigl[\sigma_j\,\varrho\,\sigma_i\, -\,
\frac{1}{2}\{\sigma_i \sigma_j\,,\,\varrho \} \Bigr]
\end{equation}
with $\sigma_i$, $i=1,2,3,$ the Pauli matrices.
The constants $C_{ij}$ come from the Fourier transform of the bath
correlation functions (see~\cite{frigerio,gor-kossak,ben-flore}) and form the so-called Kossakowski matrix
$C \equiv [C_{ij}]$.
In order to guarantee full physical consistency, namely that $\Gamma_t\otimes {\rm id}_A$ be
positivity preserving on all states of the compound system $S+A$
for any inert ancilla $A$, $\Gamma_t$ must be completely positive~\cite{lindblad1}
and this is equivalent to $C$ being positive semidefinite~\cite{gorini, lindblad2}.

In this letter we propose a way of experimentally determining the Kossakowski matrix elements,
i.e. the noise parameters,
through the transmission and reflection probabilities of an electron, which can be measured.
Moreover, in the simple case of a diagonal Kossakowski matrix, we explicitly show that, if one takes a
positive but not completely positive dissipative map, one obtains negative transmission
probabilities for certain states, which proves the necessity of complete positivity for physical consistency.

\section{Determination of the Kossakowski matrix elements}
We consider a system $S$ in which an electron propagates in a one-dimensional wire
interacting magnetically with a spin-$1/2$ impurity at $x=0$, and we study the effects of a
noisy environment on such a system.
This is the first step in a framework analogous to that in~\cite{ciccarello}.
There the authors considered an isolated system in which an electron propagates in a one-dimensional
wire interacting magnetically with two spin-$1/2$ impurities at $x=0$ and at $x=x_0$, and they analyzed the
dependence of the electron's transmittivity on the impurities' states.
Our idea was to study the effects of a noisy environment on such a system, starting by considering
the simpler system of an electron interacting magnetically with only one spin-$1/2$ impurity.
Therefore, firstly we will consider the case in which the system $S$ of the electron and one impurity is isolated;
then we will see what happens if $S$ is embedded in an external environment which
acts with a noisy term only on the spin degree of freedom of the impurity. This latter case will lead to expressions
of the Kossakowski matrix elements in terms of the electron transmission coefficients.

In the first case the eigenvalue equation for the energy is the following:
\begin{equation}\label{eigenvalue-eq-1}
H|E\rangle \equiv \left(
\frac{p^2}{2m}+\delta(x)J\vec{\sigma}\cdot\vec{s} \right) |E\rangle =
E |E\rangle \end{equation} where $p = -i\hbar \nabla$, $m$ is the electron mass,
$J$ is the magnetic coupling constant between electron and impurity, and $\vec{\sigma}$, $\vec{s}$ is the electron,
respectively impurity, spin operator\footnote{The spin operators are such that the eigenvalues of $\sigma_z$ and $s_z$
are $\pm 1/2$.}. \\
Having defined the total spin
$\begin{displaystyle} \vec{S}=\vec{\sigma} + \vec{s}
\end{displaystyle}$, we can rewrite the eigenvalue equation (\ref{eigenvalue-eq-1}) as follows:
\begin{equation}\label{eigenvalue-eq-2} \left(
\frac{p^2}{2m}+\delta(x)\frac{J}{2}\left(S^2 - \frac{3}{2}\right)
\right) |E\rangle = E |E\rangle. \end{equation}
$S^2$ and $S_z$ are the constants of motion with eigenvalues $s$ e $m=-s, \dots, s$ respectively. In
our case we are considering two spin-$1/2$ systems, so
the possible values of the total spin eigenvalues $s$ are $1$ and $0$. \\
Given an energy eigenstate $|E\rangle$, it is always possible to expand it in terms of the spatial and total spin
eigenfunctions:
$
|E\rangle=\sum_{i=1}^4 |\psi_{S_i}\rangle \otimes |S_i\rangle ,
$ where we have taken
$ \{|S_i\rangle \}_{i=1}^4 = \{ |S_1\rangle:= \frac{|01\rangle - |10\rangle}{\sqrt{2}},
|S_2\rangle:=|00\rangle, |S_3\rangle := \frac{|01\rangle + |10\rangle}{\sqrt{2}}, |S_4\rangle:=|11\rangle \}$
as the total spin basis. \\
If we project equation (\ref{eigenvalue-eq-1}) onto the electron position eigenstates $\{|x\rangle\}$,
for a fixed spin $S_i$, we get the differential equation for the wave function $\psi_{k;S_i}(x)$:
\begin{equation}\label{wave-funct-eq} -\frac{\hbar^2}{2m}
\psi''_{k;S_i}(x) + \delta(x) \frac{J}{2} \left(S_i^2 -
\frac{3}{2}\right)\psi_{k;S_i}(x) = E \psi_{k;S_i}(x).
\end{equation}
For positive energies $\begin{displaystyle}E=\frac{\hbar^2
k^2}{2m}>0\end{displaystyle}$, the solution of equation (\ref{wave-funct-eq}) is
\begin{displaymath} \psi_{k;S_i}(x) = \left\{
\begin{array}{ll}
e^{ikx} + r^E_{S_i} e^{-ikx} & \textrm{if $x<0$}\\
t^E_{S_i} e^{ikx} & \textrm{if $x>0$}
\end{array} \right.
\end{displaymath}
where $r^E_{S_i}$ and $t^E_{S_i}$ are the electron reflection and transmission coefficients respectively. \\
The explicit expressions for these coefficients are found
by imposing the continuity condition in $x=0$ and integrating the Schr\"odinger equation around $x=0$:
\begin{equation}\label{coeffs}
t^E_{S_i}=\frac{1}{1 + \frac{i}{4}\pi J \varrho(E)(S_i^2 - \frac{3}{2})} \;
\text{ , } \qquad
r^E_{S_i}=\frac{-\frac{i}{4}\pi J \varrho(E)(S_i^2 - \frac{3}{2})}{1 + \frac{i}{4}\pi J \varrho(E)(S^2_i-\frac{3}{2})} \, ,
\end{equation}
with $\varrho(E)=\frac{1}{\pi \hbar}\sqrt{\frac{2m}{E}}$ the linear density of states in the wire. \\
For the positive energies solutions, the eigenstates can be written as follows:
\begin{equation}
|E\rangle = \int dx \sum_{i=1}^4 \left[ \left( e^{ikx} + r^E_{S_{i}}
e^{-ikx} \right) \chi_L(x) + t^E_{S_{i}} e^{ikx} \chi_R(x)
\right] |x\rangle \otimes |S_i\rangle,
\end{equation}
where
$ |S_i\rangle $, $i=1,\ldots,4$, are the total spin basis elements listed above,
$r^E_{S_i}$ and $t^E_{S_i}$ are the reflection and transmission coefficients from (\ref{coeffs}),
and $\chi_{R(L)}$ is the characteristic function for $x \geq 0$ ($x \leq 0$). \\
Calculating the transmission and reflection coefficients for the spin basis elements $\{S_i \}_{i=1}^4$ explicitly,
we find
$t^E_{S_2} = t^E_{S_3} = t^E_{S_4} := t^E_1$, $t^E_{S_1} := t^E_0$, and
$r^E_{S_2} = r^E_{S_3} = r^E_{S_4} := r^E_1$, $r^E_{S_1} := r^E_0$. Thus, having redefined the spin states as
\begin{equation}\label{spin-states}
|\phi^{spin}_0\rangle :=  |S_1\rangle = \frac{|01\rangle-|10\rangle}{\sqrt{2}} , \quad
|\phi^{spin}_1\rangle := \frac{1}{\sqrt{3}}\sum_{i=2}^4 |S_i\rangle =
\frac{1}{\sqrt{3}} \left( |00\rangle +
\frac{|01\rangle+|10\rangle}{\sqrt{2}} + |11\rangle \right),
\end{equation}
the energy eigenstates can be rewritten as
$
|E\rangle = |\phi^E_0\rangle \otimes |\phi^{spin}_0\rangle +
|\phi^E_1\rangle \otimes |\phi^{spin}_1\rangle \, ,
$
where the vectors $|\phi^E_0\rangle$, $|\phi^E_1\rangle$ are such that
\begin{eqnarray}
\phi^E_0(x) \equiv \langle x|\phi^E_0\rangle & := & \left( e^{ikx} +
r^E_0 e^{-ikx}
\right) \chi_L(x) + t^E_0 e^{ikx} \chi_R(x) , \label{phi_0} \\
\phi^E_1(x) \equiv \langle x|\phi^E_1\rangle & := & \sqrt{3} \left[ \left(
e^{ikx} + r^E_1 e^{-ikx} \right) \chi_L(x) + t^E_1 e^{ikx}
\chi_R(x) \right] \label{phi_1}.
\end{eqnarray}
Finally, considering the basis of maximally entangled Bell states,
$
|\psi_0\rangle = \frac{|00\rangle + |11\rangle}{\sqrt{2}},
|\psi_1\rangle = \frac{|01\rangle + |10\rangle}{\sqrt{2}},
|\psi_2\rangle = \frac{|01\rangle - |10\rangle}{\sqrt{2}},
|\psi_3\rangle = \frac{|00\rangle - |11\rangle}{\sqrt{2}},
$
the spin states (\ref{spin-states}) can be rewritten as
\begin{equation}
|\phi^{spin}_0\rangle \equiv |\psi_2\rangle \, , \qquad \qquad
|\phi^{spin}_1\rangle \equiv \frac{1}{\sqrt{3}} \left( |\psi_1\rangle+
\sqrt{2} |\psi_0\rangle \right).
\end{equation}

If the system $S$ is embedded in an external environment to which it is weakly coupled, the evolution of the system
eigenstates is described by the Master equation (\ref{master-eq})
with solution $\varrho(t)\equiv \Gamma_t[\varrho]=\exp \left( t \mathbf{L}  \right)[\varrho] =
\exp \left( t (\mathbf{L}_H + \mathbf{L}_D)\right)[\varrho]$, where $\varrho \equiv |E\rangle \langle E|$.
We will consider a dissipative map $\Gamma_t$ whose noisy effects act only on the spin degree of freedom of the
impurity and leave the electron spin unchanged: therefore $\mathbf{L}_D \equiv \mathbb{I} \otimes \mathbf{L}_D^{(2)}$,
with $\mathbb{I}$ the identity operator acting on the space and electron spin degrees of freedom, and $\mathbf{L}_D^{(2)}$
the Kossakowski-Lindblad term (\ref{dissipation}) corresponding to the dissipative map $\gamma_t$ acting on the impurity's spin.
It is precisely the elements of the Kossakowski matrix relative to $\gamma_t$ that we want to write in terms of the
electron transmission and reflection coefficients, and thus determine operatively.
Since the total evolution of the system eigenstates is
governed by the map $\Gamma_t$ which cannot be factorized between space and spin degrees of freedom,
it is sufficient to consider an expansion for small times and stop at the first order in $t$:
\begin{equation}\label{expansion}
\Gamma_t\left[ |E\rangle \langle E| \right] = \exp \left( t \mathbf{L} \right)
\left[ |E\rangle \langle E| \right] = \left( \mathbb{I} + t \left( \mathbf{L}_H +
\mathbf{L}_D \right) \right) \left[ |E\rangle \langle E| \right] + O(t^2).
\end{equation}
Since we want to find explicit expressions for the Kossakowski matrix elements, we will isolate the contributions
from the dissipative part of the map $\Gamma_t$, due to $\mathbf{L}_D$. In order to do so, we will consider the spin state
$|\psi_3\rangle = \left(\mathbb{I}_2 \otimes \sigma_3 \right) |\psi_0\rangle$, which is orthogonal to the spin states in the
eigenstate expansion, and then we will calculate the probability of finding the system at a point $x=x_0$ with total spin state
$|\psi_3\rangle$. Therefore we will evaluate the mean value of
(\ref{expansion}) with respect to the state $|x_0\rangle \otimes |\psi_3\rangle \equiv |x_0; \psi_3\rangle$:
 \begin{equation}\label{prob-1}
P_t(x=x_0; |\psi_3\rangle \langle \psi_3|)=\langle x_0; \psi_3|\left(
\mathbb{I} + t \left( \mathbf{L}_H + \mathbf{L}_D \right) \right) \left[ |E\rangle \langle E|
\right]|x_0; \psi_3\rangle + O(t^2).
\end{equation}
The zeroth order term in (\ref{prob-1}) vanishes because of the orthogonality of $|\psi_3\rangle$ to the spin states of
the eigenstate $|E\rangle$, whereas the Hamiltonian term is always zero on the eigenstates, so we are left with
\begin{equation}\label{prob-2}
P_t(x=x_0; |\psi_3\rangle \langle\psi_3|) = t \langle x_0; \psi_3|\mathbf{L}_D \left[
|E\rangle \langle E| \right]|x_0; \psi_3\rangle + O(t^2) \, .
\end{equation}
Making use of (\ref{phi_0}) and (\ref{phi_1}), expression (\ref{prob-2}) can be conveniently rewritten as follows:
\begin{equation}\label{prob-3}
P_t(x=x_0; |\psi_3\rangle \langle \psi_3|) = t
\langle\phi^E(x)|\tilde{\emph{D}}|\phi^E(x)\rangle + O(t^2),
\end{equation}
where $\begin{displaystyle} |\phi^E(x)\rangle = \left( \begin{array}{l}
\phi^E_0(x) \\ \text{\tiny{\phantom{abc}}} \\ \phi^E_1(x)
\end{array} \right)
\end{displaystyle}$ \\ \tiny{\phantom{ancrigb}} \\
\normalsize and
\begin{equation}\label{D-tilde}
\tilde{\emph{D}} = \left(
\begin{array}{ll}
\langle\psi_3|\mathbf{L}_D\left[ |\phi_0^{spin}\rangle \langle \phi_0^{spin}|
\right]|\psi_3\rangle &
\langle\psi_3|\mathbf{L}_D\left[ |\phi_1^{spin}\rangle \langle \phi_0^{spin}| \right]|\psi_3\rangle \\
\text{\tiny{\phantom{abc}}} & \text{\tiny{\phantom{abc}}}
\\ \langle\psi_3|\mathbf{L}_D\left[ |\phi_0^{spin}\rangle \langle \phi_1^{spin}| \right]|\psi_3\rangle &
\langle\psi_3|\mathbf{L}_D\left[ |\phi_1^{spin}\rangle \langle \phi_1^{spin}|\right]|\psi_3\rangle
\end{array} \right)\,.
\end{equation}
The matrix $\tilde{\emph{D}}$ has the following explicit expression:
\begin{equation}\label{matrix_D}
\tilde{\emph{D}} = \left( \begin{array}{cc} C_{11} & \frac{1}{\sqrt{3}}\left( -iC_{21} + C_{31} \right) \\
 \frac{1}{\sqrt{3}} \left( iC_{12} + C_{13} \right)
& \frac{1}{3}\left( C_{22} + 2C_{33} + 2\sqrt{2} Im(C_{23}) \right)
\end{array} \right),
\end{equation}
where $C_{ij}$, $i,j=1,2,3$, are the elements of the Kossakowski matrix $C$ which corresponds to the map
$\gamma_t$ acting on the spin degree of freedom of the impurity. \\
In this paper we consider only entropy-increasing maps, which describe many interesting situations in
differents areas of physics (stochastic magnetic fields, quantum baker's map~\cite{alicki,soklakov,brun},
XY spin-$1/2$ chain with quenching of the transverse
field~\cite{dutta}): in this case the Kossakowski matrix $C$ is symmetric and real~\cite{lendi,breuer},
and therefore has only six different elements. So, in order to explicitly find the Kossakowski matrix elements,
we need six independent linear equations for the $C_{ij}$'s. \\
The first two linear equations are given by the explicit evaluation of (\ref{prob-3}) for $x>0$ and for $x<0$, in which
the transmission and reflection coefficients appear respectively; whereas the other four can
be obtained by calculating the analog of (\ref{prob-3}) for $x>0$ and for $x<0$ rotating the spin basis.
A simple choice for the rotations is to exchange two Pauli matrices while keeping the third fixed, thus
rearraging the elements $C_{ij}$ in~(\ref{matrix_D}). We can take a rotation $\emph{R}^{(k)}$ that keeps $\sigma_k$ ($k=1,2,3$)
fixed while changing $\sigma_l$ in $\pm\sigma_m$ ($l,m=1,2,3, \, l,m\neq k$): this will exchange the elements $C_{lm}$ and
$C_{ml}$ while leaving those with $i,j = k$ unchanged. In particular we chose
\begin{equation}\label{rotation-1}
\emph{R}^{(1)}(-\frac{\pi}{4}) = \frac{\mathbb{I}_2 - i\sigma_1}{\sqrt{2}} =
\frac{1}{\sqrt{2}}\left( \begin{array}{cc} 1 &
-i \\ -i & 1
\end{array} \right) \end{equation}
and
\begin{equation}\label{rotation-2}
\emph{R}^{(2)}(-\frac{\pi}{4}) = \frac{\mathbb{I}_2 - i\sigma_2}{\sqrt{2}} =
\frac{1}{\sqrt{2}}\left( \begin{array}{cc} 1 &
-1 \\ 1 & 1
\end{array} \right) \,
\end{equation}
which lead to the following rotated spin bases respectively:
\begin{equation}\label{rot-bases}
|\tilde{\psi}_i^{spin}\rangle := \mathbb{I}_2
\otimes \emph{R}^{(1)} |\psi_i^{spin}\rangle \qquad \text{and} \qquad
|\hat{\psi}_i^{spin}\rangle := \mathbb{I}_2
\otimes \emph{R}^{(2)} |\psi_i^{spin}\rangle .
\end{equation}
Thus we calculated (\ref{prob-3}) for $x>0$ and for $x<0$ for the three different spin bases, i.e. we evaluated
the probability of finding the evolved state at a point $x \equiv x_0$ in the three different spin states $|\psi_3\rangle$,
$|\tilde{\psi}_3\rangle$, $|\hat{\psi}_3\rangle$. We then wrote the reflection coefficient as $r_{S_i}^E=1-t_{S_i}^E$,
and finally we obtained six independent linear equations for the elements $C_{ij}$.
This linear system can be written in vector form as follows\footnote
{Details are given in the Appendix.}:
\begin{equation}\label{result_1}
|P(t)_\alpha\rangle  = M_{\alpha \beta} |C_\beta\rangle \,, \quad \quad \alpha,\beta=1,\ldots,6 \,,
\end{equation}
with $M \equiv [M_{\alpha \beta}]$ a $6\times 6$ matrix, and vectors
$$
|C_\beta\rangle :=
\begin{pmatrix}
                C_{11}  \\
                C_{12}  \\
                C_{13}   \\
                C_{22}   \\
                C_{23}   \\
                C_{33}
\end{pmatrix} \ ,\qquad \qquad
|P(t)_\alpha\rangle :=
\begin{pmatrix}
                P_0^T  \\
                P_1^T  \\
                P_2^T   \\
                P_0^R   \\
                P_1^R   \\
                P_2^R
\end{pmatrix} \, ,
$$
where $P_a^T$, $P_a^R$, $a=0,1,2$, are the transmission, respectively reflection, probabilities
for the three different spin bases. \\
The linear system of equations can be inverted if $Det(M) \neq 0$. This is indeed the case, and therefore one can
explicitly write the Kossakowski matrix elements in terms of the transmission and reflection
probabilities, and thus determine the $C_{ij}$'s experimentally. \\
Moreover, the experimental determination of the $C_{ij}$'s leads to
the possibility of actually verifying whether the Kossakowski matrix
is positive semidefinite\footnote {Notice that, in order for the
Kossakowski matrix to be positive semidefinite, the transmission and
reflection
probabilities must be such that the following positivity conditions for $C$ are fulfilled: \\
$C_{11} \geq 0, C_{22} \geq 0, C_{33} \geq 0, C_{11}C_{22} -
C_{12}^2 \geq 0, C_{11}C_{33} - C_{13}^2 \geq 0, C_{22}C_{33} -
C_{23}^2 \geq 0, Det(C) \geq 0$.}, and thus whether the evolution is
completely positive. Further, this could also be a test for the
Markovian approximation used: since physically consistent Markovian
approximations for the Master equation (\ref{master-eq}) must lead
to completely positive dynamical semigroups, if the results obtained
for the $C_{ij}$'s yield a Kossakowski matrix which is not positive
semidefinite, this could imply that the particular Markovian
approximation chosen to describe the dynamics is not appropriate.

\bigskip

\noindent \textbf{Remark 1}\quad Our proposal for an operational
determination of the Kossakowski matrix elements through
transmission probabilities, which can be measured, could also be
viewed in the context of experimental characterization of the
dynamical evolution of an open quantum system. A well-studied
procedure with this aim is known as \textit{quantum process
tomography} (QPT)~\cite{chuang, poyatos, leung}, where a quantum
system $A$ is subjected to an unknown quantum process $\mathcal{E}$.
In order to determine $\mathcal{E}$, one prepares a fixed set of
states $\{\varrho_j \}$ that form a basis for the set of operators
acting on the state space of $A$ and applies the process
$\mathcal{E}$ to each input state $\varrho_j$; then
$\mathcal{E}(\varrho_j)$ can be experimentally determined through
quantum state tomography~\cite{vogel, nielsen} on the outputs;
finally the process $\mathcal{E}$ can be fully characterized through
the operation elements $E_k$ in its operator sum representation
$\mathcal{E}(\varrho) = \sum_k E_k \varrho E_k^{\dagger}$. The
physical systems and detailed procedures used in quantum process
tomography (see, for example, ~\cite{childs, altepeter, weinstein})
differ from those used in this work, and in our case the quantum
operation describing the evolution of the system is not supposed to
be unknown; nevertheless both methods could be viewed in the context
of experimental characterization of a quantum process on an open
quantum system. Moreover, the analysis of results in quantum process
tomography leading to a non-completely positive evolution may be
useful in better understanding the implications, in our case, of
experimentally obtaining a Kossakowski matrix which is not positive
semidefinite. In~\cite{weinstein}, for example, it is shown that
experimental errors made in the QPT procedure can yield results
which lead to a non-completely positive quantum operation and that
this unphysical result can be corrected. Therefore, it might be
possible also in our case, that experimental results leading to a
non-positive semidefinite Kossakowski matrix be due to experimental
errors in the measuring procedure. Further discussion about this
hypothesis, however, would involve taking into account the exact
experimental situation, and is therefore outside the scope of this
paper.


\section{Complete positivity}
In order to guarantee full physical consistency, namely that $\gamma_t\otimes {\rm id}_A$ be
positivity preserving on all states of the compound system $S+A$ for any inert ancilla $A$,
$\gamma_t$ must be completely positive~\cite{lindblad1} and this is equivalent to $C$ being
positive semidefinite~\cite{gorini, lindblad2}. \\
The necessity of complete positivity arises from the existence of entanglement,
since in general entangled bipartite states may become non-positive under the action of positive
but not completely positive transformations~\cite{horodecki}. \\
In this section we will show that, if we take a positive but not completely positive
dissipative map $\gamma_t$ acting on the impurity's spin and calculate the probability (\ref{prob-3})
for certain entangled states, we find negative values for the transmission probability. \\
In order to give an explicit example of this fact, we will use a specific dissipative map $\gamma_t$ and
for simplicity we will consider a diagonal Kossakowski matrix.
In this case the matrix $\tilde{\emph{D}}$ from~(\ref{matrix_D}) reduces to
\begin{equation}
\tilde{\emph{D}} = \left( \begin{array}{cc} C_{11} & 0 \\ 0 &
\frac{1}{3}\left( C_{22} + 2C_{33} \right)
\end{array} \right)
\end{equation}
and the probability (\ref{prob-3}), to first order in $t$, is therefore
\begin{eqnarray}\label{prob-4}
P_t(x=x_0; |\psi_3\rangle \langle\psi_3|) &=& t\left(C_{11} |t^E_0|^2 +
(C_{22} + 2C_{33}) |t^E_1|^2 \right) \nonumber \\
&=& t \left( C_{11}
\frac{16}{16 + 9(\frac{\pi}{2} J \rho(E))^2}
\right. \left.\!+(C_{22} + 2C_{33})
\frac{16}{16 + (\frac{\pi}{2} J \rho(E))^2} \right)
\end{eqnarray}
having inserted the explicit expressions for the transmission coefficients in the second line. \\
In particular, we consider a positive but not completely positive map $\gamma_t$ acting on the impurity's spin
with Kossakowski matrix $C=$ diag(1,1,-1). Thus the Kosskowski-Lindblad term (\ref{dissipation}) explicitly
reads
\begin{equation*}
\mathbf{L}_D^{(2)}[\varrho^{spin}]= \sigma_1\varrho^{spin}\sigma_1 +
\sigma_2\varrho^{spin}\sigma_2 - \sigma_3\varrho^{spin}\sigma_3 - \varrho^{spin}
\end{equation*}
with $\varrho^{spin}$ the impurity's spin state.
Since the spin-$1/2$ impurity consists of a qubit, its state can be written in Bloch vector form:
$\varrho^{spin} = \frac{\mathbf{1}_2 + \bar{\varrho}\cdot \bar{\sigma}}{2}$, where $\mathbf{1}_2$
is the identity in $\mathbf{C}^2$, $\bar{\varrho}=(\varrho_1,\varrho_2,\varrho_3)$
and $\bar{\sigma}=(\sigma_1,\sigma_2,\sigma_3)$, with $\sigma_i$, $i=1,2,3,$ the Pauli matrices.
Thus the evolved state will be
$$\varrho^{spin}(t) \equiv \gamma_t[\varrho^{spin}] =
\frac{1 + \varrho_1\sigma_1 + \varrho_2\sigma_2 + e^{-4t}\varrho_3\sigma_3}{2} \,.$$
$\varrho^{spin}(t)$ is such that $||\varrho^{spin}(t)||^2 < ||\varrho^{spin}||^2 \leq 1$ and
therefore the map $\gamma_t$ gives rise to a positive evolution.
On the other hand, though, the Kossakowski matrix $C=$ diag(1,1,-1) is not positive semidefinite, thus the corresponding
dissipative map $\gamma_t$ is not completely positive, and evaluating (\ref{prob-4}) explicitly
it is straightforward to see that we obtain a \textit{negative} transmission probability. \\
Notice that this physical inconsistency arises from dealing with a positive but
not completely positive map and an \textit{entangled} state.
Indeed, using duality, the matrix $\tilde{\emph{D}}$ that appears in expression (\ref{prob-3}) for the
transmission probability can be rewritten as
$$ \tilde{\emph{D}} = \left(
\begin{array}{ll}\label{D-tilde}
\langle\phi_0^{spin}|\mathbf{L}_D\left[ |\psi_3\rangle \langle \psi_3|
\right]|\phi_0^{spin}\rangle &
\langle\phi_1^{spin}|\mathbf{L}_D\left[ |\psi_3\rangle \langle \psi_3| \right]|\phi_0^{spin}\rangle \\
\text{\tiny{\phantom{abc}}} & \text{\tiny{\phantom{abc}}}
\\ \langle\phi_0^{spin}|\mathbf{L}_D\left[ |\psi_3\rangle \langle \psi_3| \right]|\phi_1^{spin}\rangle &
\langle\phi_1^{spin}|\mathbf{L}_D\left[ |\psi_3\rangle \langle \psi_3| \right]|\phi_1^{spin}\rangle
\end{array} \right) \,, $$
where the spin state $|\psi_3\rangle = \frac{|00\rangle - |11\rangle}{\sqrt{2}}$ is maximally entangled for the electron and impurity spins.
Therefore, evaluating the probability (\ref{prob-4}) of finding the system at a point $x=x_0$ with total spin state
$|\psi_3\rangle$ is equivalent to applying the generator $\mathbf{L}_D \equiv \mathbb{I} \otimes \mathbf{L}_D^{(2)}$,
relative to the positive but not completely positive map $\gamma_t$, to the entangled state $|\psi_3\rangle$,
and this leads to the physical inconsistency of a negative transmission probability.

\bigskip

\noindent \textbf{Remark 2}\quad
If, instead, the map $\gamma_t$ is completely positive, expression (\ref{prob-3}) is always positive.
Indeed, from the choice of the spin state
$|\psi_3\rangle$, the only contribution to (\ref{prob-2}) is given by the noise term $N[\cdot]$,
which can be written as follows:
\begin{eqnarray*}
N[\varrho] &=& \sum_{i,j}C_{ij}\sigma_j\varrho\sigma_i^\dagger =
\sum_l c_l\Big( \sum_j\bar{\psi}_l^{(j)}\sigma_j\Big)\varrho\Big( \sum_i\psi_l^{(i)}\sigma_i^\dagger\Big) \\
&=& \sum_l c_l\Big( \sum_j\bar{\psi}_l^{(j)}\sigma_j\Big)\varrho\Big( \sum_i\bar{\psi}_l^{(i)}\sigma_i\Big)^\dagger
= \sum_l \sqrt{c_l}\tilde{W}_l\varrho\sqrt{c_l}\tilde{W}_l^\dagger \equiv \sum_l W_l \varrho W_l^\dagger \,,
\end{eqnarray*}
where $\{|\psi_l\rangle\}$ is a basis of eigenstates of the Kossakowski matrix
$C \equiv \sum_l c_l |\psi_l\rangle \langle \psi_l|$ such that $C_{ij}=\sum_l c_l \psi_l^{(i)}\bar{\psi}_l^{(j)}$.
$N[\varrho]$ is in Kraus-Stinespring form and thus completely positive.
Therefore, rewriting the action of the dissipative generator $\mathbf{L}_D$ in (\ref{prob-2}), we have:
$$
P_t(x=x_0;|\psi_3\rangle \langle \psi_3|) = t\langle x_0;\psi_3|\mathbb{I} \otimes N^{spin}
[|E\rangle \langle E|]|x_0;\psi_3\rangle + O(t^2),
$$
where $\mathbb{I}$ is the identity operator for the position and electron spin degrees of freedom,
while $N^{spin}$ is the noise term acting on the impurity's spin. \\
The complete positivity of the latter guarantees the
positivity of $\mathbb{I} \otimes N^{spin}$, and thus
$P_t(x=x_0;|\psi_3\rangle \langle \psi_3|)$ is always positive. \\

\section{Conclusions}
In this paper we proposed a way of experimentally determining the
elements $C_{ij}$ of a Kossakowski matrix, i.e. the noise
parameters, in terms of the transmission and reflection
probabilities of an electron, which can be measured. We considered a
system in which an electron propagates in a one-dimensional wire
interacting magnetically with a spin-$1/2$ impurity, embedded in an
external environment which acts with a noisy term only on the spin
degree of freedom of the impurity. We calculated the electron's
transmission and reflection coefficients, and found expressions for
the transmission and reflection probabilities in terms of the
Kossakowski matrix elements. This leads to the possibility of having
experimental access to the noise parameters and of actually
verifying whether the Kossakowski matrix is positive semidefinite,
and thus whether the evolution is completely positive. Further, it
could also be a test for the Markovian approximation used: if the
results obtained for the $C_{ij}$'s lead to a Kossakowski matrix
which is not positive semidefinite, this could imply that the
particular
Markovian approximation used to describe the dynamics is not appropriate.  \\
Moreover, we gave a concrete example of the necessity of complete positivity for physical consistency,
showing that a positive but not completely positive dissipative map acting on the impurity's spin can yield negative
transmission probabilities for certain entangled states.
\newpage

\noindent \textbf{Appendix}\quad
The first two equations for the transmission and reflection probabilities in terms of the Kossakowski matrix elements
and transmission coefficients, computed to the first order in $t$, are as follows:
\begin{eqnarray*}
\frac{P_0^T(t)}{t} &=& C_{11}|t_0^E|^2 + C_{12}\big[-2Im(t_0^E)Re(t_1^E)+2Im(t_1^E)Re(t_0^E)\big]\\
&+& C_{13}\big[2Re(t_0^E)Re(t_1^E)+2Im(t_1^E)Im(t_0^E)\big]+
C_{22}|t_1^E|^2 +2C_{33}|t_1^E|^2 \, ,
\\ \phantom{abchgt} \\
\frac{P_0^R(t)}{t} &=& C_{11}\big[2-2\cos(2kx)+|t_0^E|^2 -2Re(t_0)+2Re(t_0)\cos(2kx)+2Im(t_0)\sin(2kx)\big] \\
&+& C_{12}\big[-2Im(t_0^E)Re(t_1^E)+2Im(t_1^E)Re(t_0^E) +2Im(t_0^E) - 2Im(t_1^E) \\ &+& 2Re(t_0)\sin(2kx) -Im(t_0)\cos(2kx)-
2Re(t_1)\sin(2kx)+Im(t_1)\cos(2kx)\big] \\
&+& C_{13}\big[4-4\cos(2kx)+2Re(t_0^E)Re(t_1^E)+2Im(t_1^E)Im(t_0^E)+2Re(t_1)\cos(2kx) \\ &+& 2Im(t_1)\sin(2kx)-2Re(t_1)-2Re(t_0)+2Re(t_0)\cos(2kx)
+2Im(t_0)\sin(2kx)\big] \\
&+& C_{22}\big[2-2\cos(2kx)+|t_1^E|^2-2Re(t_1)+2Re(t_1)\cos(2kx)+2Im(t_1)\sin(2kx)\big] \\
&+& 2C_{33}\big[2-2\cos(2kx)+|t_1^E|^2-2Re(t_1)+2Re(t_1)\cos(2kx)+2Im(t_1)\sin(2kx)\big] \, .
\end{eqnarray*}
The other four equations for $P_a^{T,R}$, $a=1,2$, have the same form just with a different ordering of the elements $C_{ij}$
in accordance with the rotation of the spin bases, as previously explained, using (\ref{rotation-1}), (\ref{rotation-2}), (\ref{rot-bases}).
In particular, taking $x=\frac{n\pi}{4k}$ with $n=4m+1$ and $m=0,1,2,\ldots$, we get $\sin(2kx)=1$, $\cos(2kx)=0$:
this leads to six distinct equations where the expressions for the
reflection probabilities are somewhat simplified whereas those for the transmission probabilities remain unchanged.
If we then write this linear system in vector form, we get (\ref{result_1}),
where $M$ is a matrix whose entries are the constants multiplying the elements $C_{ij}$ in the system of equations.
This matrix explicitly reads:
$$
M=
\begin{pmatrix}
                a_0  &   b  &  c  &  a_1  &  0 & 2a_1 \\

                a_0  &  -c  &  b  &  2a_1 & 0  & a_1 \\

                2a_1 &  0  & -c  &  a_1  &  b  & a_0 \\

                d_0 &  e  &  f  &  d_1  &  0  &  2d_1  \\

                d_0 &  -f  &  e &  2d_1  & 0  & d_1 \\

                2d_1  & 0  &  -f  & d_1  &  e  &  d_0
\end{pmatrix} \ ,
$$
with
\begin{eqnarray*}
&& a_i=Re(t_i^E)=|t_i^E|^2 \,, \qquad \qquad i=0,1 \,,  \\
&& b=2[-Im(t_0^E)Re(t_1^E)+Im(t_1^E)Re(t_0^E)], \\
&& c=2[Re(t_0^E)Re(t_1^E)+Im(t_1^E)Im(t_0^E)],  \\
&& d_i=2-|t_i^E|^2+2Im(t_i^E) \,, \qquad \qquad i=0,1 \,,  \\
&& e=2[Im(t_0^E)Re(t_1^E)+Im(t_1^E)Re(t_0^E)+Re(t_0^E)-Re(t_1^E)+Im(t_0^E)-Im(t_1^E)],  \\
&& f=2[2+ Re(t_0^E)Re(t_1^E)+Im(t_1^E)Im(t_0^E)-Re(t_0^E)-Re(t_1^E)+Im(t_0^E)+Im(t_1^E)].
\end{eqnarray*}


\noindent \textbf{Aknowledgement}\quad
The authors would like to thank F. Benatti for his valuable comments and suggestions,
and R. Floreanini for his help in reviewing the manuscript.


\begin{thebibliography}{99}

\bibitem{lendi} R. Alicki, K. Lendi,
{\it{Lecture Notes in Physics}}, vol. 286, Springer-Verlag (1987)

\bibitem{verri}
V. Gorini, A. Frigerio, M. Verri, A. Kossakowski, E. C. G. Sudarshan, \textit{Rep. Math.
Phys.} {\bf{13}}, 149 (1976)

\bibitem{spohn}
H. Spohn, {\it{Rev. Mod. Phys.}},
 {\bf{52}}, 569 (1980)

\bibitem{kossak1}
A. Kossakowski, \textit{Rep. Math. Phys.} \textbf{3}, 247 (1972)

\bibitem{kossak2}
A. Kossakowski, \textit{Bullettin de l'Academie Polonaise des
Sciences} \textbf{20}, 12 (1972)

\bibitem{gorini}
V. Gorini, A. Kossakowski, E. C. G. Sudarshan, \textit{J. Math.
Phys.} {\bf{17}}, 821 (1976)

\bibitem{dumcke}
R. D\"umcke, H. Spohn, \textit{Z. Phys.},
\textbf{B34}, 419 (1979)


\bibitem{ben-flore} F. Benatti, R. Floreanini,
\textit{International Journal of Modern Physics B} \textbf{19}, 19
(2005)


\bibitem{frigerio}
A. Frigerio, V. Gorini, \textit{J. Math.
Phys.} {\bf{17}}, 2123 (1976)

\bibitem{gor-kossak}
V. Gorini, A. Kossakowski, \textit{J. Math.
Phys.} {\bf{17}}, 1298 (1976)


\bibitem{lindblad1} G. Lindblad,
\textit{Comm. Math. Phys.} {\bf{40}}, 147 (1975)

\bibitem{lindblad2} G. Lindblad,
\textit{Comm. Math. Phys.} {\bf{48}}, 119 (1976)

\bibitem{ciccarello}
F. Ciccarello, G. M. Palma, M. Zarcone, Y. Omar, V. R. Vieira,
\textit{New J. Phys.},
\textbf{8}, 214 (2006)


\bibitem{alicki}
R. Alicki, A. Lozinski, P. Pakonski, K. Zyczkowski, \textit{J. Phys. A},
 \textbf{37}, 5157 (2004)


\bibitem{soklakov}
A. N. Soklakov, R. Schack, {\it{Phys. Rev. E}},
 {\bf{66}}, 036212 (2002)

\bibitem{brun}
T. A. Brun, R. Schack, {\it{Phys. Rev. A}},
 {\bf{59}}, 2649 (1999)


\bibitem{dutta}
V. Mukherjee, A. Dutta, D. Sen, \textit{Phys. Rev. B} \textbf{77}, 214427 (2008)

\bibitem{breuer} H. P. Breuer, F. Petruccione,
{\it{The theory of Open Quantum Systems}}, Oxford University Press (2002)


\bibitem{chuang}
I. L. Chuang, M. A. Nielsen,
 {\it{J. Mod. Opt.}} {\bf{44}}, 2455 (1997)

\bibitem{poyatos}
J. F. Poyatos, J. I. Cirac, P. Zoller,
 {\it{Phys. Rev. Lett.}} {\bf{78}}, 390 (1997)

\bibitem{leung}
D. W. Leung, \textit{Towards Robust Quantum Computation},
Ph.D. thesis, Stanford University (2000)

\bibitem{vogel}
K. Vogel, H. Risken,
 {\it{Phys. Rev. A}} {\bf{40}}, 2847 (1989)

\bibitem{nielsen}
M. A. Nielsen, I. L. Chuang,
 {\it{Quantum Computation and Quantum Information}}
 (Cambridge UP, Cambridge, 2000)

\bibitem{childs}
A. M. Childs, I. L. Chuang, D. W. Leung, {\it{Phys. Rev. A}}
{\bf{64}}, 012314 (2001)

\bibitem{altepeter}
J. B. Altepeter et al.,
 {\it{Phys. Rev. Lett.}} {\bf{90}}, 193601 (2003)

\bibitem{weinstein}
Y. S. Weinstein et al., {\it{J. Chem. Phys.}} {\bf{121(13)}},
6117-6133 (2004)

\bibitem{horodecki}
M. Horodecki, P. Horodecki, R. Horodecki, {\it{Phys. Lett. A}} \textbf{223}, 1 (1996)

\end{thebibliography}
\end{document}